# Plasmonic Band-stop MIM Waveguide Filter Based on Bilateral Asymmetric Equilateral Triangular Ring

Yang Gao(高扬)[1]*, Jincheng Wang(王金成)[1], Hengli Feng(冯恒利)[1], Jingyu Zhang(张景煜)[1], Zuoxin Zhang(张作鑫)[1], Dongchao Fang(房冬超)[1], Chang Liu(刘畅)[1], Lehui Wang(王乐慧)[1]

[1]School of Electronic Engineering, Heilongjiang University, Harbin, 150080, China
[2]Heilongjiang Provincial Key Laboratory of Metamaterials Physics and Device
*Corresponding author: gaoy_hit@163.com

In this paper, a bilateral asymmetric equilateral triangular ring (ETR) band-stop filter based on metal-insulator-metal (MIM) waveguide is proposed. The transmission spectrum and electric field distribution of the filter are simulated and theoretically investigated by finite difference time domain (FDTD) and coupled mode theory (CMT) method. The results show that changing the number and position of resonant cavities can adjust the transmission characteristics of the filter. The minimum transmission and sensitivity of this filter is 0.19% and 1149 nm/RIU, respectively. The filters have broad prospects in the highly integrated optical circuits and refractive index sensor.
**Keywords**: *Surface plasmon polaritons; Metal-insulator-metal; Waveguide;*
DOI: 10.3788/COLXXXXXX.XXXXXX.

## 1. Introduction

Surface plasmon polaritons (SPPs) are the electromagnetic waves propagating at the interface between metal and insulator layers with different signs of dielectric constants. It is usually produced at the interface between the medium and the negative dielectric constant material[1]. Since it enables the aggregation of light energy on the sub-wavelength scale, so the amplitude of electromagnetic energy is evanescent wave in the vertical direction of the interface[2-5]. SPPs has remarkable properties such as overcoming the diffraction limit and manipulating light at a sub-wavelength scale[6-10]. Therefore, different types of SPPs-based devices such as filters, couplers, refractive index sensors[11-15], Bragg reflectors[16,17], Mach-Zehnder interferometers[18], etc. have been numerically simulated and experimentally demonstrated. In addition, plasmonic waveguides have received extensive attention due to their simple structures and easy fabrication. There are two basic types of SPPs waveguides, insulator-metal-insulator (IMI) waveguide and metal-insulator-metal (MIM) waveguide[19, 20]. The IMI waveguide mainly concentrates on the order of microns, and the loss is small in long-distance transmission, but it can not limit the light in sub-wavelength scale, so it is not suitable for high optical integration. The MIM waveguide has strong confinement of light and acceptable propagation length for SPPs, so they are widely used in subwavelength devices.

In recent years, MIM surface plasmon waveguide filters with different structures have also been widely studied, such as single-sided multi-tooth waveguide filters[21], double-ring and rectangular ring filters[22], side-coupled beam square ring resonators[23], etc. In this paper, a new ETR filter with high filtering efficiency is designed. The simulated and theoretically investigated of ETR filters were studied by finite difference time domain (FDTD) and coupled mode theory (CMT)[24-27]. The results show that the minimum transmission of the filter is 0.19%. Besides, the sensitivity of the ETR filter reaches 1149 nm/RIU, which is higher than previous reports. The filter has broad application prospect in highly integrated optical circuits and refractive index sensors.

## 2. Structural model and theoretical analysis

The schematic diagram of the proposed waveguide structure is shown in Fig. 1(a), which includes an equilateral triangular ring and straight waveguide (Filter 1). The finite difference time domain (FDTD) method is employed to simulate and study its optical properties, and mesh step is set as $\Delta x = \Delta y = 5$ nm. In Fig. 1 (a), the white part represents air, and its relative dielectric constant $\varepsilon_a = 1$, and the blue part represents silver. The relative permittivity of silver is expressed by Drude model[28]:

$$\varepsilon_s(\omega) = \varepsilon_\infty - \frac{\omega_p^2}{\omega(\omega + i\gamma)}, \quad (1)$$

where $\varepsilon_\infty$ is the dielectric constant at infinite frequency, $\omega$ is the angular frequency of incident light, $\omega_p$ is the plasma resonance frequency, representing the natural frequency of free conductive electron oscillation, $\gamma$ is the electron collision frequency. The parameters for silver can be set as $\varepsilon_\infty = 3.7$, $\omega_p = 1.38 \times 10^{16}$ rad/s, and $\gamma = 2.37 \times 10^{13}$ rad/s.

In Fig. 1(a), the distances of the two triangular bottom edges from the straight waveguide are $h_1$ and $h_2$, respectively. The side lengths of the triangular are $l_1$ and $l_2$, respectively. The waveguide width is $W$. The specific parameters of the waveguide structure are shown in Table 1.

**Table 1.** Parameters of the designed filter.

| Parameter | $h_1$ | $h_2$ | $l_1$ | $l_2$ | $W$ |
|---|---|---|---|---|---|
| Value (nm) | 238 | 270 | 300 | 200 | 50 |

Because the width of the straight waveguide in the structure is much smaller than the wavelength of the incident light, the excited SPPs will propagate along the interface of metal. This waveguide structure only supports TM wave. According to the boundary continuity condition of electromagnetic field, the chromatic dispersion relation of the TM polarized surface plasmons in MIM waveguide can be derived[29,30].

$$\tanh\left(\frac{wk_0\sqrt{n_{eff}^2 - \varepsilon_a}}{2}\right) = -\frac{\varepsilon_a\sqrt{n_{eff}^2 - \varepsilon_s}}{\varepsilon_s\sqrt{n_{eff}^2 - \varepsilon_a}}, \quad (2)$$

$$n_{eff} = \frac{\beta_{spp}}{k_0} = \frac{\beta_{spp}\lambda}{2\pi}, \quad (3)$$

$$k_0 = \frac{2\pi}{\lambda}, \quad (4)$$

where $k_0$ is the wave number in the vacuum, $n_{eff}$ is the effective refractive index of waveguide structure, $\beta_{spp}$ is the propagation constant of SPPs, $\lambda$ is the wavelength of incident light in free space.

When the incident light propagates in the straight waveguide, it can be coupled into the resonant cavity and form a standing wave, so that some specific wavelengths of light can not pass through the waveguide. Resonant wavelength can be obtained by the standing wave theory[31,32]: $\lambda = L_{eff} \operatorname{Re}(n_{eff})/m$ ($m$ =1, 2). In the formula, $L_{eff}$ is the effective length of the ring resonant structure, $n_{eff}$ is the effective refractive index of the resonant cavity, $m$ is the resonant order, ($m$ represents the first and second order resonance, respectively). Because SPPs decay exponentially in the direction perpendicular to the interface, when the resonance is generated, the electromagnetic field energy is mostly concentrated inside the resonant cavity.

Next, we introduce the coupled mode theory (CMT) to explore the coupling effects and the spectral response behaviors of the system. The transmission spectrum derived from CMT and FDTD are shown in Fig. 1, where three resonance peaks are observed. They are $\lambda_1$ =517nm, $\lambda_2$ =640nm and $\lambda_3$ =1195nm, respectively. In Fig. 1 (a), $S_{n\pm}$ ($n$=1, 2, 3, 4) denotes the input and output energies, subscript $\pm$ denotes the direction of energy transfer, and $a_n$ ($n$=1, 2) denotes the energy amplitude of the $n$th resonance mode.

$$\frac{da_1}{dt} = (-i\omega_1 - \frac{1}{\tau_{w1}} - \frac{1}{\tau_{i1}})a_1 + S_{1+}\sqrt{\frac{1}{\tau_{w1}}} + S_{2-}\sqrt{\frac{1}{\tau_{w1}}}, \quad (5)$$

$$\frac{da_2}{dt} = (-i\omega_2 - \frac{1}{\tau_{w2}} - \frac{1}{\tau_{i2}})a_2 + S_{3+}\sqrt{\frac{1}{\tau_{w2}}} + S_{4-}\sqrt{\frac{1}{\tau_{w2}}}, \quad (6)$$

where $\omega_n$ ($n$=1, 2) are the resonant angular frequencies in the three resonant modes, $1/\tau_{wn} = \omega_n/2Q_{wn}$ are the decay rate due to energy transfer in the waveguide, $1/\tau_{in} = \omega_n/2Q_{in}$ are the intrinsic loss in the resonant modes. $Q_{wn}$, $Q_{in}$ and $Q_{cn}$ represent the external loss quality factor of the $n$th resonant mode, the internal loss quality factor, and the quality factor of the loss between different resonant modes, respectively. The quality factor in resonant mode is $Q_n = \lambda_n/FWHM$, and these quality factors satisfy the following relationships: $1/Q_n = 1/Q_{in} + 1/Q_{cn}$.

When the incident light propagates in the waveguide, a phase shift occurs after coupling with the resonant cavity. Based on the principle of conservation of energy, the following relationship can be obtained that:

$$S_{2+} = S_{1+} - \sqrt{\frac{1}{\tau_{w1}}}a_1, \quad S_{1-} = S_{2-} - \sqrt{\frac{1}{\tau_{w1}}}a_1, \quad (7)$$

$$S_{4+} = S_{3+} - \sqrt{\frac{1}{\tau_{w2}}}a_2 - \sqrt{\frac{1}{\tau_{w3}}}a_3, \quad S_{3-} = S_{4-} - \sqrt{\frac{1}{\tau_{w2}}}a_2 - \sqrt{\frac{1}{\tau_{w3}}}a_3, \quad (8)$$

$$S_{3+} = S_{1+}e^{i\varphi}, \quad S_{4-} = S_{2-}e^{i\varphi}, \quad (9)$$

Using the boundary condition $S_{2-}$ =0 and Eqs. (7)-(9), the amplitude transmission of the system can be derived as follows:

$$t = \frac{S_{2+}}{S_{1-}} = e^{i\varphi}(1 + \frac{\frac{\gamma_1}{\tau_{w1}} + \frac{\gamma_2}{\tau_{w2}} + \sqrt{\frac{\alpha^2}{\tau_{i2}\tau_{w2}}}}{\gamma_1\gamma_2 - \alpha^2})Ce^{i(\varphi_1+\varphi_2)} + \frac{\sqrt{\frac{\alpha^2}{\tau_{i2}\tau_{w2}}}}{\gamma_1\gamma_2 - \alpha^2}, \quad (10)$$

$$\gamma_n = i(\omega - \omega_n) - \frac{1}{\tau_{wn}} - \frac{1}{\tau_{in}} \quad (n=1, 2, 3), \quad (11)$$

$$\alpha = \sqrt{\frac{C^2 e^{2i(\varphi_1+\varphi_2)}}{\tau_{w1}\tau_{w2}}}, \quad (12)$$

Finally, the transmission can be derived as: $T = \frac{P_{out}}{P_{in}} = |t|^2$.

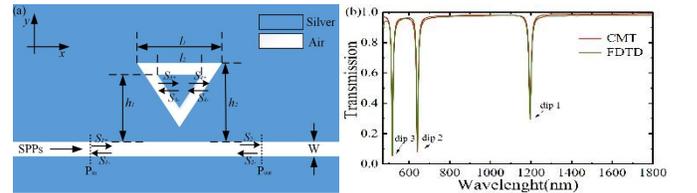

Fig. 1(a) Schematic diagram of input and output energy of filter 1. (b) Transmission characteristic curves from single-sided ETR FDTD and CMT.

## 3. Simulation Results and Discussion

### 3.1. Single-sided ETR MIM waveguide

As shown in Fig. 1(b), there are three resonance peaks which are $\lambda_1$ =1195nm, $\lambda_2$ =640nm and $\lambda_3$ =517nm. The minimum transmission of three resonant modes are 29.6%, 9.6%, 5.9% and the Q-factors are $Q_1$ =99, $Q_2$ =69, $Q_3$ =62, respectively. The corresponding magnetic field diagrams are shown in Figs. 2(b)-2(d), respectively, and it is obvious that when the incident light in the straight waveguide is coupled into the resonant cavity, the energy is concentrated in the ETR resonant cavity, forming a standing wave, and only a small amount of light can pass through the straight waveguide.

In Fig. 2(a), when $h_1$ and $h_2$ are reduced from 268 nm, 300 nm to 238 nm, 270 nm, respectively, the filtering effect of filter 1 is continuously enhanced. Since the distance between the resonant cavity and the straight waveguide needs to be considered, $h_1$, $h_2$ are 238nm, 270nm, respectively.

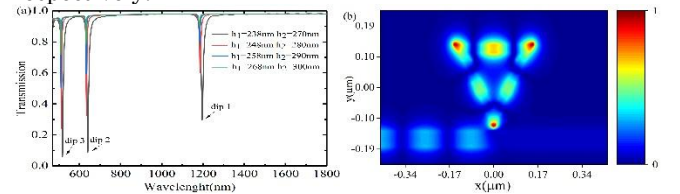

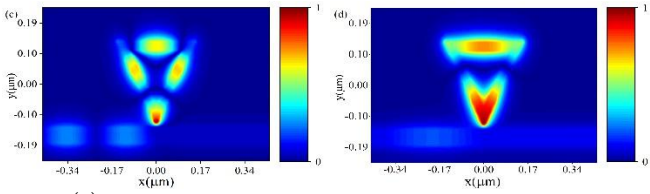

Fig. 2(a) Transmission spectrum of filter 1 with different h1, h2. (b) Magnetic field diagram of the filter 1 at 517 nm. (c) Magnetic field diagram of the filter 1 at 640 nm. (d) Magnetic field diagram of the filter 1 at 1195 nm.

### 3.2 Symmetrical ETR type MIM waveguide

Based on the structure of filter 1, a resonant cavity is added on the other side of the straight waveguide, as shown in Fig. 3(a). Compared with filter 1, the filtering effect of filter 2 is obviously enhanced. Fig. 3(b) display a comparison of the transmission spectrum of filter 2 and filter 1. Where the resonant wavelength of each peak was not shifted. The Q-factors of the three peaks are $Q_1$ =68, $Q_2$ =43, and $Q_3$ =37, respectively. In Fig. 3(b), the minimum transmission rate of filter 2 is 15.6%, 3.6%, 2.1%, respectively. Figs. 3(c)-3(e) displays the magnetic field diagram of the filter 2 structure. In the figure, when the incident light passes through the straight waveguide structure, the energy is no longer concentrated only in the upper resonant cavity. Since filter 2 is a symmetrical structure, the energy couples relatively equally to both sides of the straight waveguide as it passes through the straight waveguide. In this case, filter 2 has a smaller transmission rate than filter 1.

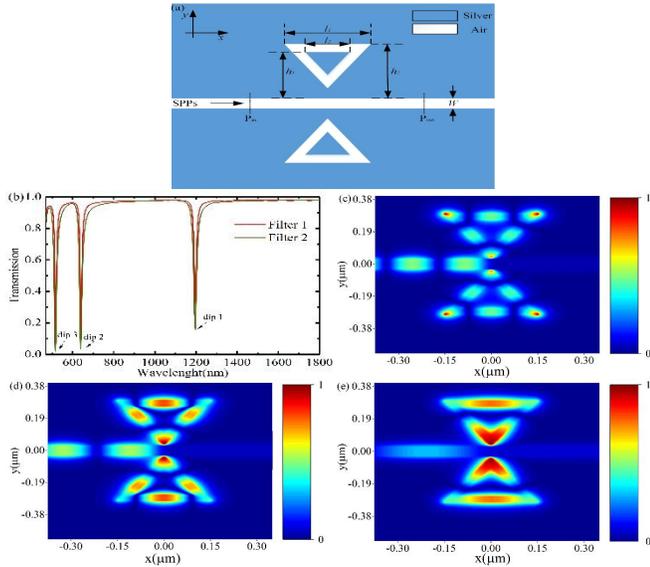

Fig. 3(a) Schematic diagram of filter 2 structure. (b) Comparison of transmission curves of filter 1 and filter 2. (c) Magnetic field diagram of the filter 2 at 517 nm. (d) Magnetic field diagram of the filter 2 at 640 nm. (e) Magnetic field diagram of the filter 2 at 1195 nm.

### 3.3 Asymmetric ETR type MIM waveguide

In the analysis of filter 2, it is found that there is a very obvious enhancement of the filtering effect after adding an identical structure to the lower side of the straight waveguide. In order to find the structure with the best performance. We continue to improve the structure of filter 2 to produce better results. A schematic diagram of the structure of the asymmetric ETR filter (filter 3) is shown in Fig. 4(a). Fig. 4(b) display the transmission spectrum obtained by varying the distance $L$ between the two resonant cavity rings. Where Q-factors are $Q_1$ =77, $Q_2$ =52, and $Q_3$ =45, respectively. As the increasing $L$, the transmission of filter 3 is enhanced. When $L$ = 100 nm, the transmission of the three resonance peaks was enhanced to 9.4%, 0.5%, 0.1%, respectively. In Figs. 4(c)-(e), the magnetic fields corresponding to the three resonant wavelengths are shown. As the filter becomes an asymmetric structure, part of the energy is first coupled to the upper side resonant cavity, and then a small amount of energy is coupled to the lower side resonant cavity. This leads to a reduction in the minimum transmission.

We also study the sensitivity characteristics of this structure to the environment. Fig. 4(f) shows the transmission spectrum of different refractive indices $n$ under this structure. It can be found that as the refractive index increases, the transmission spectrum shows obvious red shift. Here the sensitivity is defined as: $S = \Delta\lambda / \Delta n$, where $\lambda$ and $n$ are the refractive index change and the corresponding wavelength shift of the resonance peak, respectively. Where the sensitivity of filter 3 can be 1149nm/RIU. And it is obvious from Fig. 4(f) that the minimum transmission of the three resonance peaks is linearly related to the variation of the refractive index. Therefore, the present structure can be used as a sensor for monitoring environmental refractive index, ambient temperature, and liquid concentration. The structure is simpler, easier to manufacture and more efficient.

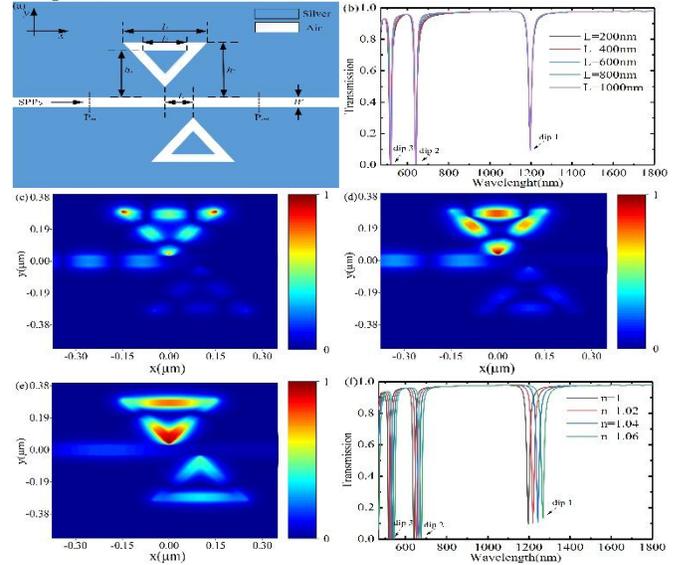

Fig. 4(a) Schematic diagram of filter 3 structure. (b) Transmission curves of filter 3 at different lengths L. (c) Magnetic field diagram of the filter 3 at 517 nm. (d) Magnetic field diagram of the filter 3 at 640 nm. (e) Magnetic field diagram of the filter 3 at 1195 nm. (f) Transmission curves of filter 3 with different refractive indices.

In order to show the results of the three filters more clearly, the resonance wavelengths and minimum transmittance of the three structures are summarized in Table 2. As shown in Table 2, it is concluded that the

resonant wavelengths of the three filters do not change when adding resonant
cavities or changing the position of resonant cavities. However, the minimum transmission keeps decreasing.

Table 2 Comparisons between the proposed filters.

| Filter | Resonance wavelength | Minimum Transmission | Resonance wavelength | Minimum Transmission | Resonance wavelength | Minimum Transmission |
|---|---|---|---|---|---|---|
| Filter 1 | 1195 nm | 0.296 | 640 nm | 0.096 | 517 nm | 0.059 |
| Filter 2 | 1195 nm | 0.156 | 640 nm | 0.036 | 517 nm | 0.021 |
| Filter 3 | 1195 nm | 0.094 | 640 nm | 0.005 | 517 nm | 0.001 |

## 4. Conclusion

In this paper, a bilateral asymmetric ETR band-stop filter based on MIM waveguide is designed, simulated and theoretically investigated. The results demonstrate that changing the number and position of resonant cavities can adjust the transmission characteristics of the filter, but it does not change the resonant wavelength of the filter. The minimum transmission and sensitivity of this filter is 0.19% and 1149 nm/RIU, respectively. According to the mentioned properties, this structure can have a wide range of applications in the highly integrated optical circuits and refractive index sensor.

## Acknowledgments

This research was funded by Natural Science Foundation of Heilongjiang Province (Grant No. LH2019F047). Project of The Central Government Supporting The Reform and Development of Local Colleges and Universities (Grant No. 2020YQ01), Heilongjiang University Outstanding Youth Science Fund of Heilongjiang University (JCL201404), Heilongjiang University Youth Science Fund Project (QL201301) and The Science and Technology Research Project of the Education Department of Heilongjiang Province (12541633).